\documentclass[preprint]{aastex631}

\shorttitle{4U 1820-30}
\shortauthors{Chou \& Jhang}

\begin{document}

\title{Updated Orbital Ephemeris and Detection of Superhump Modulation in X-ray Band for the Ultra-Compact Low Mass X-ray Binary 4U 1820-30}

\correspondingauthor{Yi Chou}
\email{yichou@astro.ncu.edu.tw}

\author[0000-0002-8584-2092]{Yi Chou}
\affiliation{Graduate Institute of Astronomy, National Central University \\
  300 Jhongda Rd. Jhongli Dist. Tauyuan, 32001, Taiwan}

\author{Yao-Wun Jhang}
\affiliation{Graduate Institute of Astronomy, National Central University \\
  300 Jhongda Rd. Jhongli Dist. Tauyuan, 32001, Taiwan}



\begin{abstract}

  The 4U 1820-30 is a ultra-compact low mass X-ray binary (LMXB) near the center of the globular cluster NGC 6624. Its negative orbital period derivative, observed from the phase evolution of its sinusoidal-like orbital variation, contradicts  the positive value obtained from the theoretical prediction. In this paper, we present the analysis of the 4U 1820-30 orbital modulation from  light curves obtained from the {\it Neutron star Interior Composition ExploreR (NICER)} observations from 2017 to mid 2022. Combined with historical records, the orbital derivative is measured from the orbital phase evolution between 1976 and 2002 is $\dot P /P =(-5.21 \pm 0.13) \times 10^{-8}$ yr$^{-1}$. No significant second order orbital period derivative is detected with a 2$\sigma$ upper limit of $|\ddot P|<5.48 \times 10^{-22}$ s s$^{-2}$. We discuss the possible intrinsic orbital period derivative of 4U 1820-30 and suggest that this binary system may have a significant mass outflow similar to some other LMXBs. In addition, a periodic modulation with a period of $691.6 \pm 0.7$ s, which is consistent with the superhump period discovered in the far ultraviolet band of the {\it Hubble Space Telescope}, was also detected in in the X-ray light curves collected by {\it NICER}. We conclude that such modulation is probably caused by a period of $0.8 \pm 0.1$ day apsidal precession of accretion disk similar to the SU UMa type dwarf novae and some LMXBs. However we cannot exclude the possibility that it is induced by a hierarchical third star orbiting around the binary system.

\end{abstract}



\section{Introduction} \label{sec:intro}

The ultra-compact low mass X-ray binary (LMXB) 4U 1820-30, located near the center of the globular cluster NGC 6624, was discovered by~\citet{gia74}. The detection of its Type I X-ray burst~\citep{gri76} indicated that the accretor of this accreting system is a neutron star. ~\citet{ste87} first reported its sinusoidal-like orbital modulation in X-ray band with a period of 685.0118 s. This orbital period is the shortest among all known LMXBs making 4U 1820-30 the most compact X-ray binary whose mass-losing companion is a Roche-lobe filling helium white dwarf with a mass of 0.06-0.08 $M_{\sun}$~\citep{rap87}. ~\citet{pri84} discovered a long-term variation in 4U 1820-30 luminosity by a factor of $\sim$3 with a period of of $\sim$176 days detected from the observations of the {\it Vela 5B} spacecraft. In addition to revising the period to be 171.003$\pm$0.326 days,~\citet{chou01} found that this period is stable for over $\sim$30 years with a period change rate $|\dot P_{long}/P_{long}| < 2.2 \times 10^{-4}$ $yr^{-1}$ and proposed that 4U 1820-30 could be a hierarchical triple with a third star orbiting the binary system with a period of $\sim$1.1 days, inducing binary orbital eccentricity modulations with a period of $\sim$171 days~\citep{maz79,gri88}. This long-term periodicity was further confirmed by~\citet{sim03}.

In addition to the orbital modulation observed in the X-ray band,~\citet{and97} discovered a modulation with a period of 687$\pm$2.4 s in the ultraviolet (UV) band from the {\it Hubble Space Telescope (HST)}. Although this period is slightly longer than the orbital period detected in the X-ray band, the difference is not sufficiently significant to distinguish between them. On the other hand, from the far-ultraviolet (FUV) light curve collected by {\it HST},~\citet{wan10} reported the discovery of a period of 693.5$\pm$1.3 s modulation, whose period is only $\sim$1\% longer but significantly different from the orbital period detected in the X-ray band. They suggested that this period is superhump period although it might be consistent with the hierarchical triple system model.

Since the discovery of orbital modulation by~\citet{ste87}, several studies were conduct to refine the orbital period and constrain the orbital period derivative from historical and newly observed data, such as~\citet{sma87},~\citet{mor88} and~\citet{sam89}. The first significant orbital derivative of 4U 1820-30 was detected by~\citet{tan91} with a value of $\dot P/P = (-1.08 \pm 0.19) \times 10^{-7}$ $yr^{-1}$. This value was further refined by~\citet{van93a,van93b} and ~\citet{chou01}. After combining all the observations in the X-ray band,~\citet{peu14} reported the latest value of orbital period derivative as $\dot P/P = (-5.3 \pm 0.3) \times 10^{-8}$ $yr^{-1}$ using the orbital modulation phase observed from 1976 to 2010. All the aforementioned detected values indicate that the orbital period derivative is negative, which contradicts to the theoretical predicted lower limit ($\dot P/P > 8.8 \times 10^{-8}$ $yr^{-1}$) proposed by~\citet{rap87}, who considered 4U 1820-30 to be an accreting system that consists of a neutron star and a mass-losing helium white dwarf. Even when considering other effects, such as tidal dissipation in the white dwarf driven by the eccentricity of the binary system, the intrinsic orbital derivative may be lower but should still be positive~\citep{pro12}.  

To explain the contradiction between the observed (negative) and theoretically predicted (positive) values of the orbital derivative of 4U 1820-30, several scenarios have been proposed.~\citet{tan91} first suggested that this deviation might be caused by the binary system accelerating in the gravitational potential of globular cluster, which was supported by~\citet{chou01}. However, with consideration of the projected location of 4U 1820-30 in the globular cluster, distance of NGC 6624 and mass distribution evaluated by the mass-to-light ratio across the cluster, the maximum acceleration from the gravitational potential of the globular cluster ($a_{max}/c \approx 10^{-9} yr^{-1}$) is approximately an order of magnitude smaller than the one that is required to explain the disparity between theoretical and observed orbital period derivative values ($a/c \approx 10^{-8} yr^{-1}$)~\citep{pro12,peu14}.~\citet{peu14} discussed some other possibilities got further accelerating the binary system, including stellar mass dark remnant, intermediate-mass black hole and central concentration of dark remnants. To preserve the stability of the triple system, which is considered the cause of 171 days of long-term modulation~\citep{chou01}, as well as the abnormal period derivative of three pulsars (J1823-3021A, B, C) in NGC 6624, the central concentration of dark remnants is the most likely scenario to explain the discrepancy. On the other hand,~\citet{jia17} suggested that the negative orbital period derivative observed in 4U 1820-30 is possibly produced by the temporary circumbinary disk which can effectively extracted the orbital angular momentum from the binary system through resonant interaction.

In this paper, we present our analysis results for the orbital modulation of 4U 1820-30 using the {\it Neutron star Interior Composition ExploreR (NICER)} observations from 2017 to 2022. Combined with historical records, the orbital phase evolution of $\sim$46 years enable us not only to further refine the orbital period but also measure or constrain the second order orbital derivative ($\ddot P$)in order to verify the possible scenarios for the negative orbital derivative observed in 4U 1820-30. In section~\ref{od}, we describe the {\it NICER} observations of 4U 1820-30, as well as data screening and grouping. The data analysis results are demonstrated in section~\ref{da}, including the update of the orbital period derivative ($\dot P/P$), constraint on the second order orbital derivative ($\ddot P$), and  detection of a period of $691.6 \pm 0.7$ s superhump modulation in X-ray light curve. In section~\ref{obcr}, we discuss the possible intrinsic orbital period derivative and suggest that the observed orbital period derivative of 4U 1820-30 is improper for inferring the gravitational acceleration in NGC 6624. Finally, the superhump and hierarchical triple interpretations for the  period of $691.6 \pm 0.7$ s modulation are also demonstrated in Section~\ref{sm}.


\section{Observations and Data Reduction} \label{od}
{\it NICER}~\citep{gen17}, aboard the International Space Station, is a soft X-ray telescope designed to investigate neutron stars through X-ray timing. Its X-ray Timing Instrument (XIT) is composed 56 X-ray concentrators (XRCs) and associated Focal Plane Modules (FPMs). The 56 FPMs are further divided into seven groups, each group consisting of eight FPMs, controlled by a Management and Power Unit (MPU). There are four FPMs, including detector ID 11, 20, 22 and 60, which have been inactive since launched as indicated in the {\it NICER Mission Guide}\footnote{$https://heasarc.gsfc.nasa.gov/docs/nicer/mission\_guide/$}.

\begin{figure}
\plotone{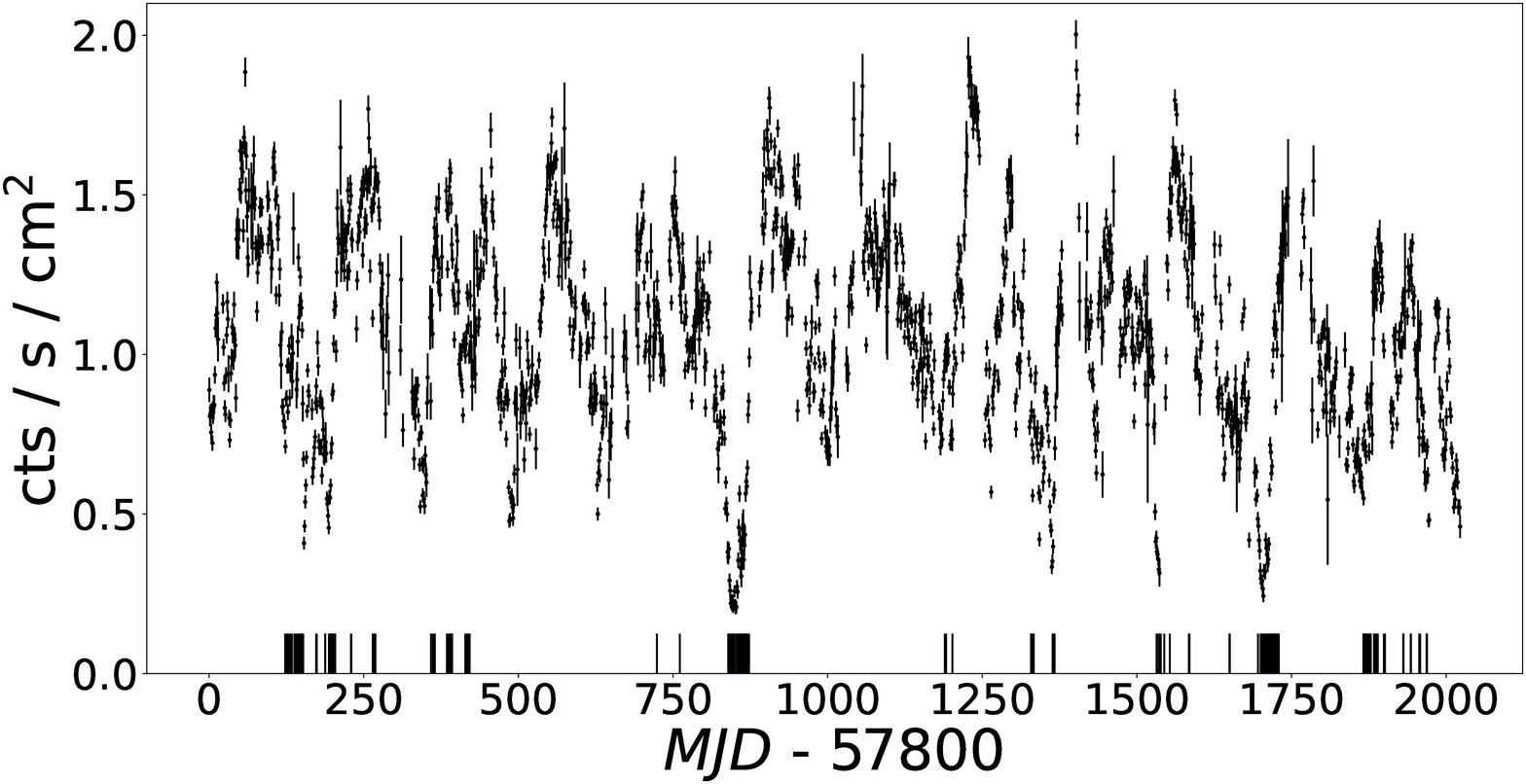}
\caption{ {\it NICER} observation distribution of 4U 1820-30. The light curve is detected by {\it MAXI} and the black bars represent the {\it NICER} observation times. \label{nicerobs}}  
\end{figure}

{\it NICER} made 173 observations of 4U 1820-30 between June 26 2017 and July 9 2022 (Figure~\ref{nicerobs}). All data files analyzed in this study were reduced using NICERDAS version 9, as a part of HEASOFT v6.30. The events were filtered using the standard screening criteria described in the {\it NICER Mission Guide}. Their photon arrival times were corrected to the barycenter of the solar system using the tool {\it barycorr} by applying the JPL solar planetary ephemeris DE430. However, the barycenter correction could not be processed for observations with ObsID 0050300118, 1050300103, 2050300119, 5604010304 and 5604010305 because the orbital files did not match the corresponding data\footnote{See $https://heasarc.gsfc.nasa.gov/docs/nicer/data_analysis/workshops/NICER-Workshop-QA-2021.pdf$}, therefore these observations were removed from the analysis. To reduce the noise and background, only the events with both slow and fast chains triggered and an energy range between 1 and 10 keV (i.e. PI values between 100 and 1000) were selected for further analysis.

\begin{figure}
\plotone{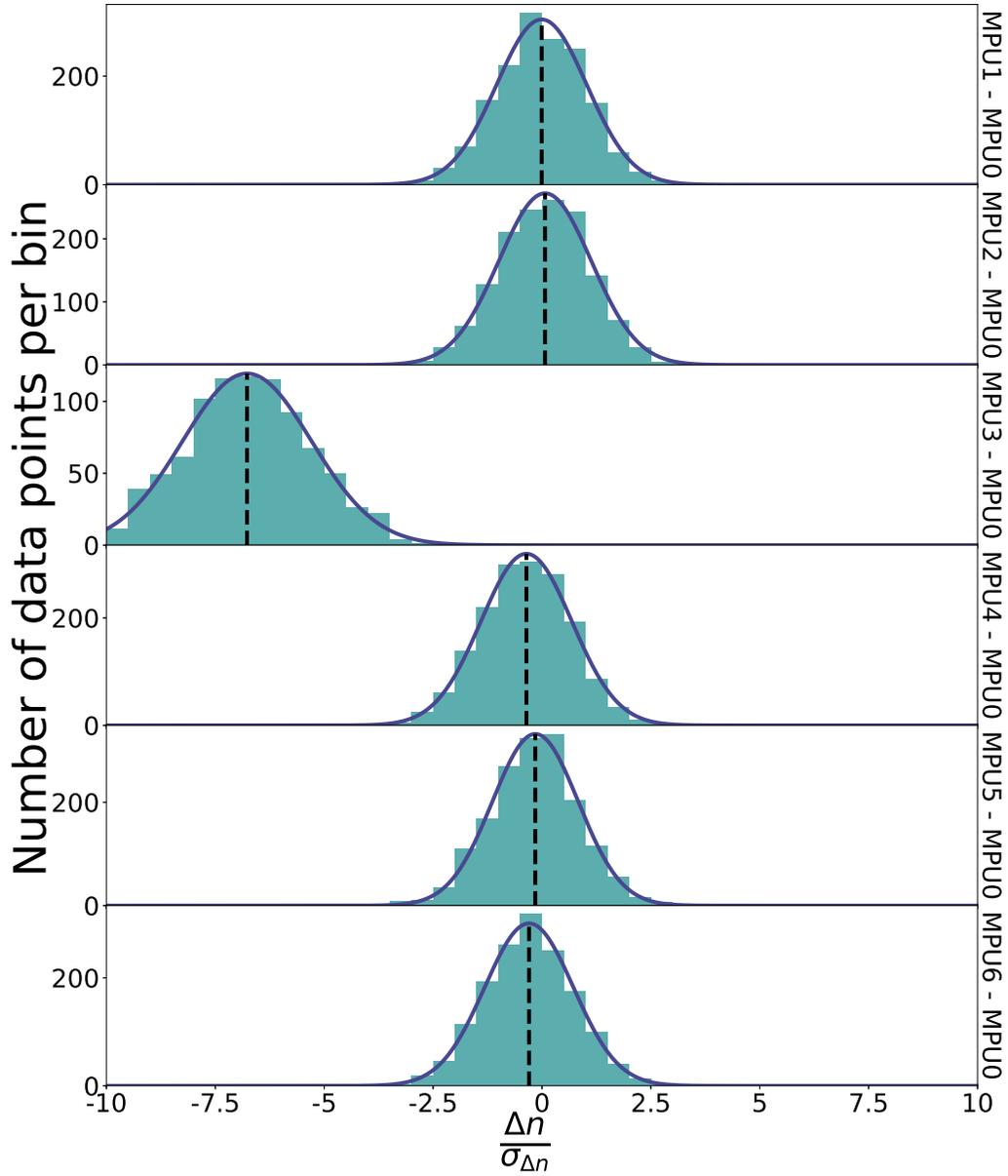}
\caption{Distribution of $\Delta n_i/\sigma_{\Delta n_i}$ for the observation with ObsId 1050300126. All centroids (dashed lines) are close to the expected value ($<\Delta n_i>=0$) except $\Delta n_3$, which indicates that MPU3 has significant lower exposure time than the others. \label{low_exp}}  
\end{figure}

However, we found that the number of active FPMs was less than 52 after data screening in certain observations, for example, only 47 active FPMs for the observation with ObsID 4680010126. Therefore, for the light curves, we normalized the count rate to the number of active FPMs, i.e., cts/s/FPM, in this work. Furthermore, some observations showed significantly lower exposure times in an MPU (mostly in MPU3) than in others. Hence, we  performed the following process to identify low-exposure MPU. First, all  events in an observation were binned into 1 s bin where $N_i$ is the number of the events of the ith MPU in this bin. $n_i$ is defined as the $N_i$ being normalized by the number of active FPMs in this MPU, i.e., $n_{i} \equiv N_{i}/fpm_{i}$ where $fmp_i$ is the number of active FPMs in the ith MPU and the uncertainty of $n_i$ can be written as $\sigma_i= \sqrt{N_i}/fpm_i$. Taking MPU0 as a reference because the low exposure has never been observed in this MPU, we defined deviation as $\Delta n_i \equiv n_{i}-n_{0}$ thus, the corresponding uncertainty is $\sigma_{\Delta n_{i}}=\sqrt{\sigma_{i}^2+\sigma_{0}^2}$. We expected that $<\Delta n_i>=0$ if the exposure time of the ith MPU was the same as that for MPU0. Figure~\ref{low_exp} shows the distributions of $\Delta n_{i}$ for an observation in which the centroid of $\Delta n_{3}$ is significantly lower than the others. Therefore, we considered the mean deviation larger than 1$\sigma$ as the low exposure time and the events in this MPU were removed from the analysis. Finally, several Type-I X-ray bursts were detected in all observations. Burst events, detected 100 s after the beginning of the bursts, were also excluded from analysis.


Because the orbital variation of 4U 1820-30 in the X-ray band is rather weak (~3\% peak-to-peak), to significantly detect the modulation profile, a light curve with sufficient long exposure time is essential. However, the exposure times of many observations were too short to detect the orbital modulation. Therefore we attempted to combine consecutive observations with short exposure times to create a group to make the light curve. The total exposure time of each group was at least 2000 s with the total time span of less than six days to minimize the trend due to long-term (171 days) variation. All other sporadic observations with short exposure times were excluded from the analysis. Table~\ref{nicergroup} shows the grouping of observations for the analysis described in the following section.



\section{Data Analysis} \label{da}
\subsection {Orbital Period Derivative} \label{opd}
\begin{figure}
\plotone{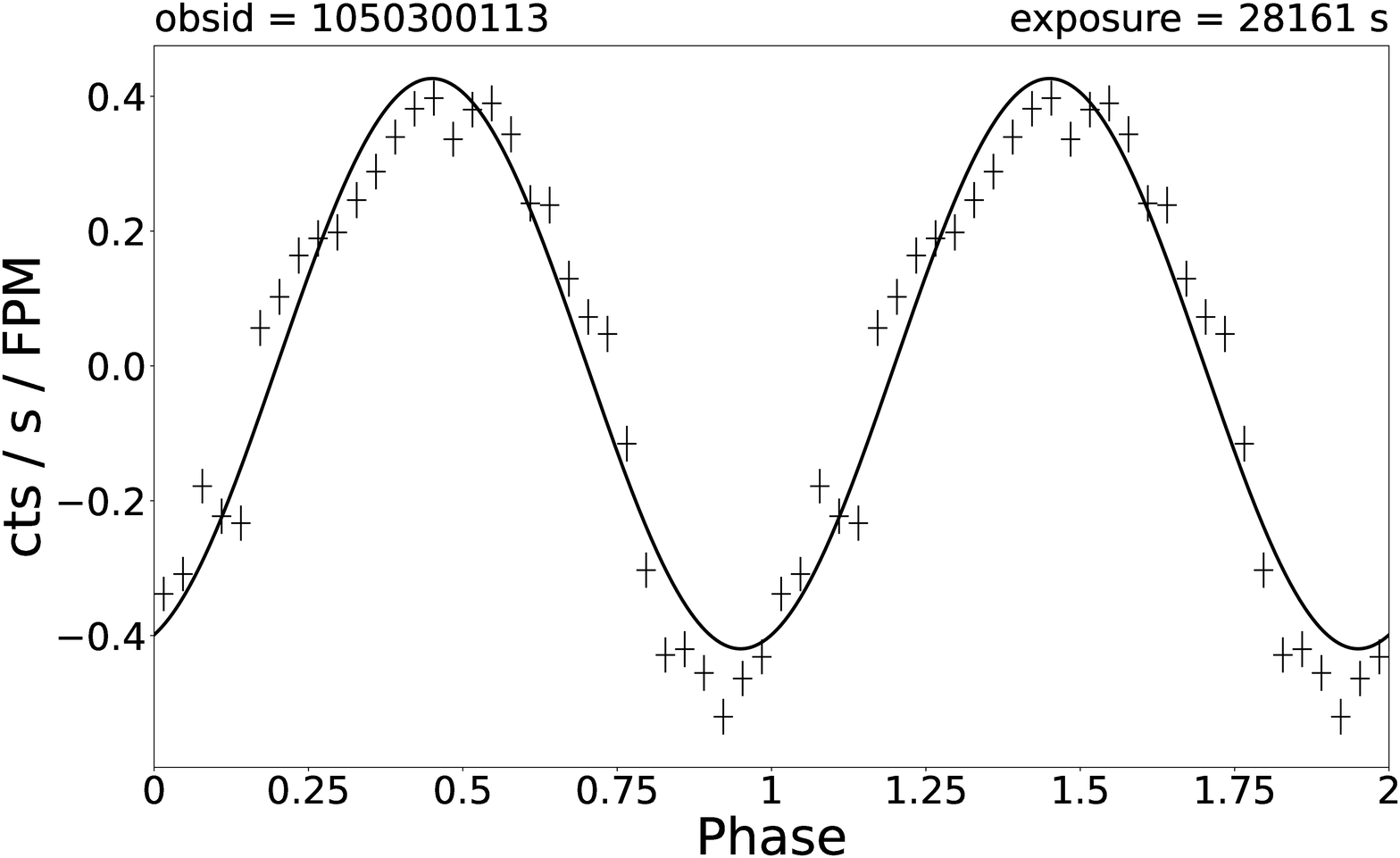}
\caption{Typical orbital modulation profile observed on September 3 2017 (ObsID 1050200113) folded by the linear ephemeris as shown in Eq.~\ref{li-eph}. The solid line is the best sinusoidal fitting for the modulation\label{fold_lc}}
\end{figure}

The events selected for each group were binned to a light curve with a bin size of 1 s. Because there were ~171 days of long-term modulation, to correctly obtain the orbital modulation profile, we removed the trend by fitting a polynomial up to 4th order for each the light curve. The de-trended light curves were folded using the revised linear ephemeris proposed by~\citet{peu14}.

\begin{equation}\label{li-eph}
T_{N} ={BJD_{TDB}2442803.63600}+{\big({{685.0118} \over {86400}}\big)} \times N
\end{equation}
\noindent where $T_{N}$ is the time of the maximum intensity of the Nth cycle and $N$ is the cycle count. However, a small number of the folded light curves showed insignificant orbital modulation. We applied analysis of variance (AoV)~\citep{sch89} to filter them out. In AoV, the variable $\Theta_{AoV}$ is defined as 

\begin{eqnarray}\label{aov}
  \nonumber  (r-1)s_1^2 &=& \sum_{i=1}^r n_i ({\bar x}_i-{\bar x})^2\\
  \nonumber  (n-r)s_2^2 &=& \sum_{i=1}^r \sum_{j=1}^{n_i} (x_{ij}-{\bar x}_{i})^2\\
  \Theta_{AoV} &\equiv&  s_1^2 / s_2^2
 \end{eqnarray}

\noindent where $n$ is the total number of the data points in the light curve, $\bar x$ is the average count rate of the light curve, $r$ is the number of bins in the folded light curve, $n_i$ is the number of data points in the ith bin of the folded light curve, ${\bar x}_{i}$ is the average count rate in the ith bin of the folded light curve and $x_{ij}$ is the jth count rate of the light curve in the ith bin of the folded light curve. If the light curve is pure Gaussian white noise, $ \Theta_{AoV}$ follows the F-distribution with $r-1$ and $n-r$ degrees of freedom. We selected $r=32$ for the testing process. Only the folded light curves with false alert probability less than 0.0053 (approximately more than 3$\sigma$ significance) were accepted for further analysis. Each folded light curves with significant orbital modulation detection was then fitted by a single sinusoidal function and the maximum value of the fitted curve was taken as fiducial point for the orbital phase. A typical orbital modulation profile is shown in Figure~\ref{fold_lc}. In a manner similar to that done by~\citet{chou01} and ~\citet{peu14}, we combined the orbital phases of each year using weighted averaging. The phase jitter, as seen in {\it Rossi X-ray Timing Explorer (RXTE)} data~\citet{chou01}, can be observed in the phases of each year. Therefore, we estimated the phase jitter value for each year based on the standard deviation of the phases of the year. The final error for the phase of each year was evaluated by quadratically adding the phase jitter and average phase error from the fitting of the folded light curve. The mean arrival time of the average phase for each year was calculated using the Equation 2 in~\citet{chou01}. The average orbital phases measured from 2017 to 2022 {\it NICER} observations are listed in Table~\ref{nicerphase}.

\begin{figure}
\plotone{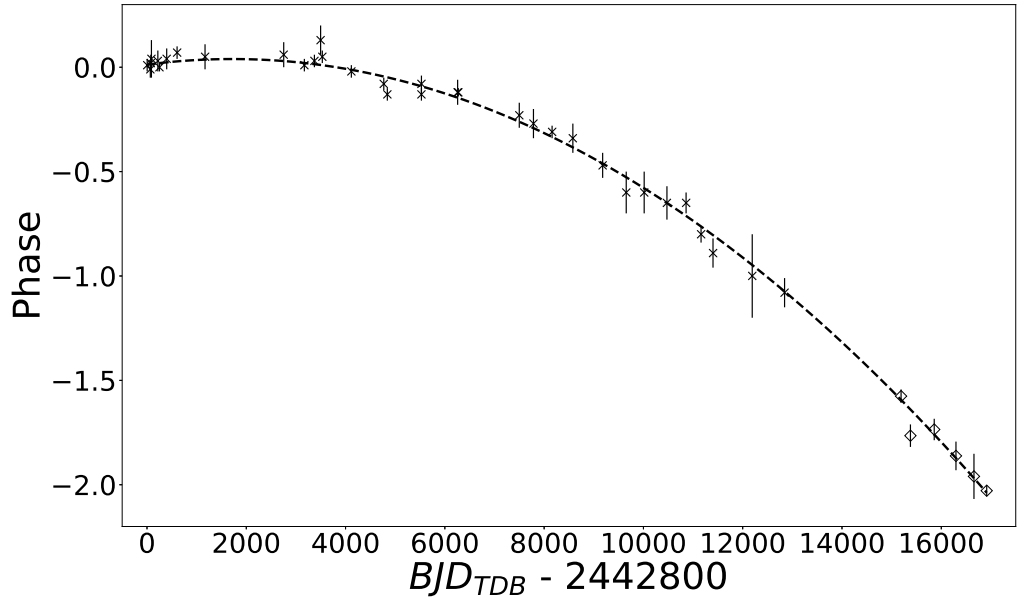}
\caption{Orbital phase evolution of 4U 1820-30 from 1976 to 2022 folded by the linear ephemeris shown in Eq~\ref{li-eph}. The filled diamond are the orbital phases from {\it NICER} observations and the dashed line is the optimal quadratic fit result.\label{qua_fit}}
\end{figure}

In combination with the historical records of the orbital phases listed in Table 1 of~\citet{peu14}, the evolution of the orbital phases of 4U 1820-30 measured between 1976 and 2022 are shown in Figure~\ref{qua_fit}. Given that a significant orbital period derivative has been detected~\citep[$> 17 \sigma$,][]{peu14}, we directly fitted the orbital phase evolution with the quadratic function shown in Eq~\ref{qua_fit_eq} assuming  that the orbital period derivative ($\dot P$) was a constant over the entire time span of $\sim 46$ years   

\begin{equation}\label{qua_fit_eq}
\Phi(t)={{T_0-T_{fold}} \over P_{fold}} +{\Big({{P_0 - P_{fold}} \over {P_0 P_{fold}}  } \Big)}t+ {1 \over 2}{\dot P \over {{P_0}}^2} t^2
\end{equation}
\noindent where $P_0$ is the orbital period at the expected phase zero epoch $T_0$, $T_{fold}=BJD_{TDB}2442803.63600$, $P_{fold}=685.0118$ s and $t=T_{mean} - T_0$ where $T_{mean}$ is mean arrival time of the average phase of each year. We obtained a reasonable fitting with $\chi^2/\nu=33.72/37$. This fitting results give an orbital period derivative of $\dot P=(-1.131 \pm 0.029) \times 10^{-12}$ s s$^{-1}$, or $\dot P /P =(-5.21 \pm 0.13) \times 10^{-8}$ yr$^{-1}$, which is almost identical to those found by~\citet{van93b} and~\citet{peu14} but with smaller uncertainty. The quadratic ephemeris can be updated as follows:

\begin{equation}\label{qua_eph}
T_N = {(BJD_{TDB}2442803.636101 \pm 0.00094)}+{\big({{685.011968 \pm 0.000021} \over {86400}}\big)} \times N + (4.48 \pm 0.12) \times 10^{-15} \times N^2
\end{equation}

\noindent We also attempted to probe the second order orbital period derivative ($\ddot P$) suggested by~\citet{peu14}. We fitted the phase evolution with a cubic function and obtained a second order orbital period derivative of $\ddot P=(9.5 \pm 27.4) \times 10^{-23}$ s s$^{-2}$ with $\chi^2/\nu=33.60/36$. In comparison with quadratic model, the F-test gave an F-value of 0.13 and a p-value of 0.72, which implied that the cubic model improved the fitting with confidence level only 28\%. We therefore conclude that there is no significant second order orbital period that can be detected with a $2\sigma$ upper limit of $|\ddot P|<5.48 \times 10^{-22}$ s s$^{-2}$

\subsection{Search for Superhump Modulation in X-ray Light Curves} \label{smx}


\citet{wan10} discovered a periodic modulation in the FUV light curves of 4U 1820-30 with a period of 693.5 s, using data collected by the HST. They suggested that this modulation, which has a period about 1.2\% longer than the orbital period, is the superhump period, based on similarities between 4U 1820-30 and the dipping LMXB X 1916-053. In X 1916-053, the superhump period is not only seen in the optical band but also in X-ray light curves~\citep{chou01a}. For 4U 1820-30, the sinusoidal-like orbital modulation in the X-ray band is believed to be caused by absorption of structures in the disk rim, similar to X-ray dipping sources~\citep{ste87}. Therefore, it is likely that the superhump modulation is detectable in the X-ray light curves of 4U 1820-30.

\begin{figure}
\plotone{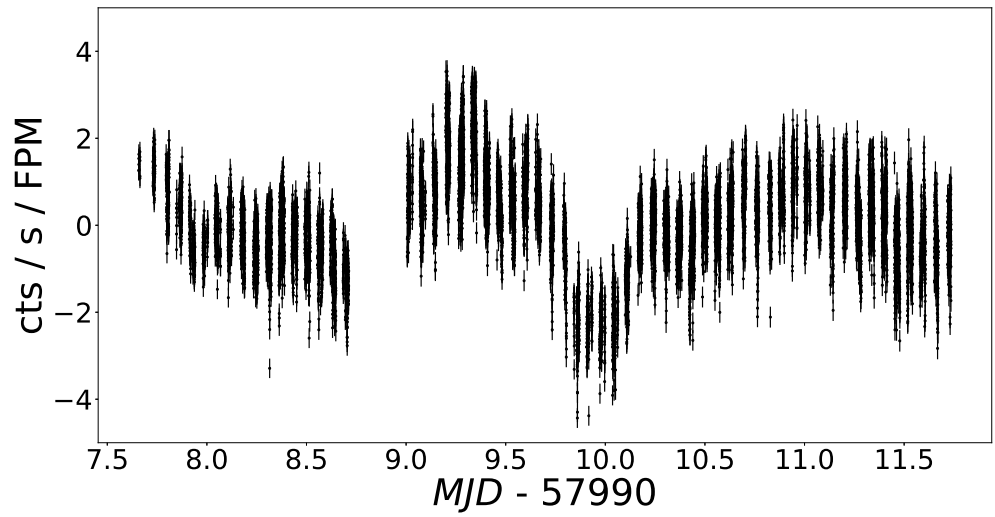}
\caption{Trend removed light curve collected by {\it NICER} between September 1 and 5, 2017 for the superhump modulation search.\label{sup_lc}}
\end{figure}

\begin{figure}
\plotone{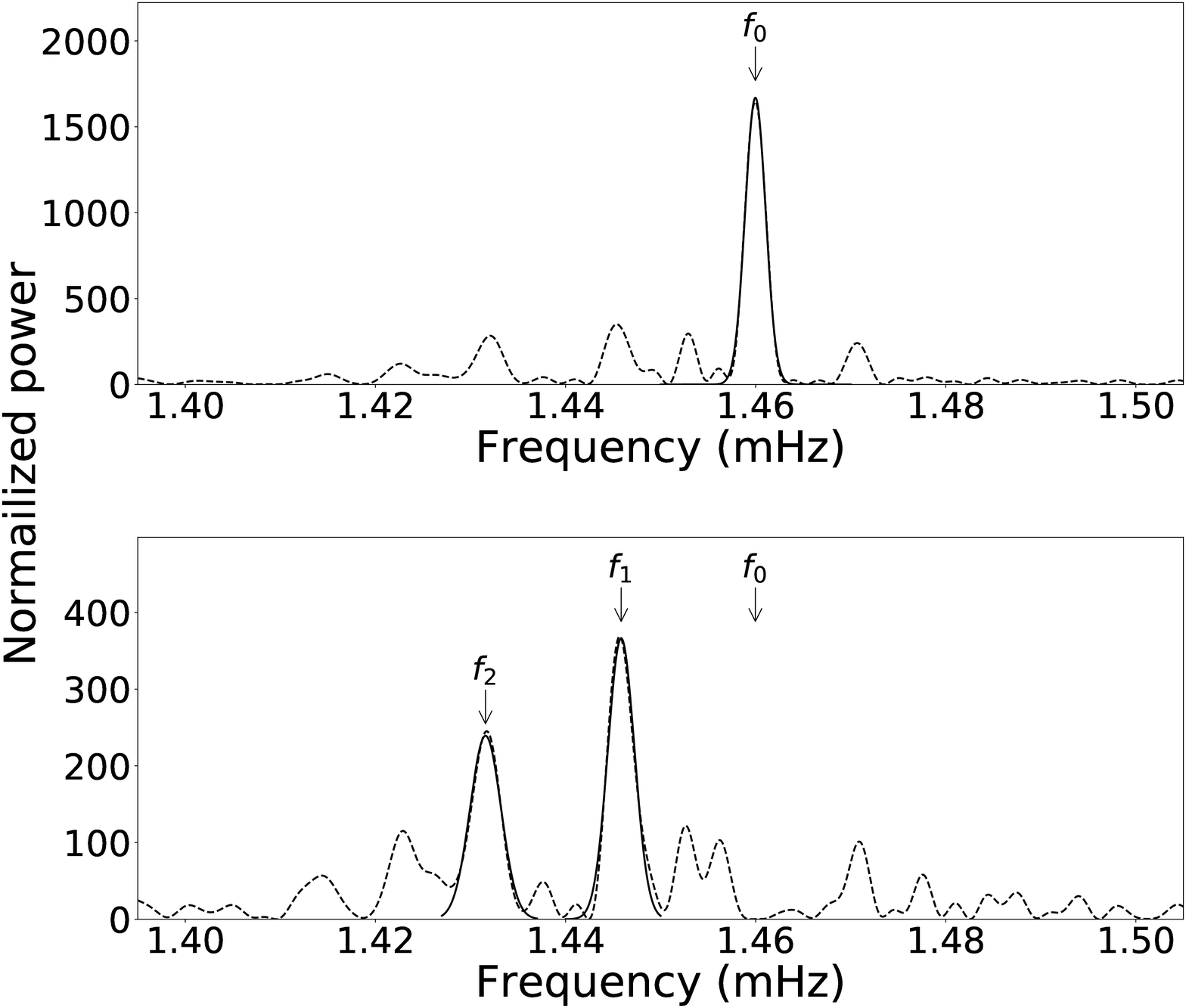}
\caption{Top panel: power spectrum (dashed line) of  trend removed light curve collected by {\it NICER} between September 1 and 5, 2017. The main peak is from the orbital modulation. A Gaussian function fitting (solid line) resulted in a frequency $f_0 = 1.460 \pm 0.001$ mHz (or $P_0 = 684.9 \pm 0.5$ s). Bottom panel: power spectrum (dashed line) of residual light curve (after the orbital modulation being removed). The frequencies of the two peaks are $f_1 = 1.446 \pm 0.001$ mHz (or $P_1 = 691.6 \pm 0.7$ s) and $f_2 = 1.432 \pm 0.002$ mHz (or $P_2 =698.3 \pm 0.8$ s) evaluated by Gaussian fittings (solid lines). The $f_1$ is consistent with the superhump frequency proposed by \citet{wan10}.  \label{sup_powspec}}
\end{figure}

\begin{figure}
\plotone{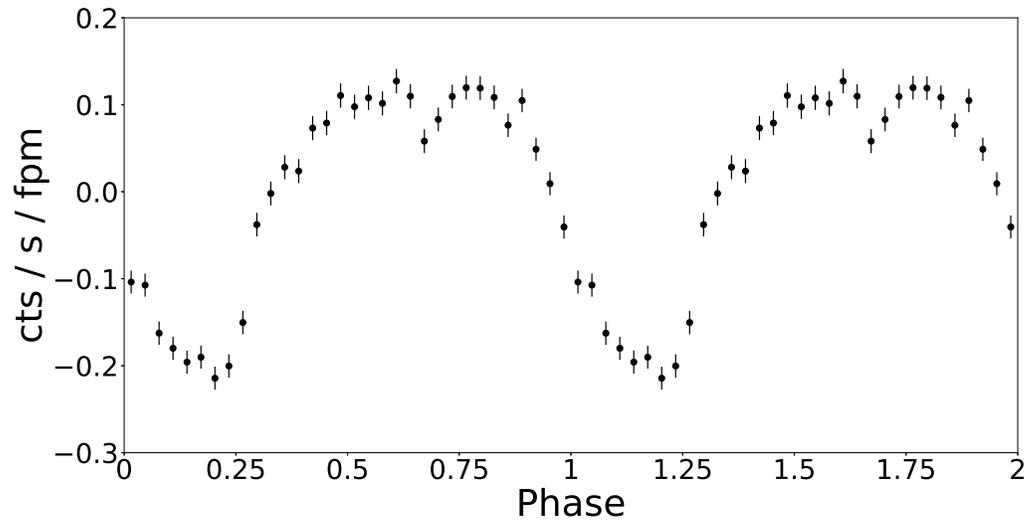}
\caption{Residual light curve folded by the 691.6 s superhump period detected in this work with an arbitrary phase zero epoch.\label{sup_mod}}
\end{figure}

To resolve the superhump modulation signal ($\sim$693 s) from the orbital variation ($\sim$685 s) in the power spectrum, as well as to avoid too much alias induced by the observation gaps, a continuous observation with a time span much more than $\sim$0.8 day was preferred. We found that the {\it NICER} observations made between September 1 to 5 2017 with ObsId from 1050300111 to 1050300115 were suitable for searching for superhump modulation. The trend-removed light curve, with a total exposure time of $\sim$104 ks and time span of $\sim$4 days is shown in Figure~\ref{sup_lc}. To search for the periodicity of this 1 s binned light curve, we adopted the Lomb-Scargle (LS) periodogram~\citep{sca82} to obtain the power spectrum as shown in Figure~\ref{sup_powspec}. The most significant peak is located at $\sim$1.46 mHz, which is caused by the orbital modulation. Similar to the analysis performed by~\citet{wan10}, we fitted a Gaussian function to the main peak and obtained the peak frequency $f_0=1.460 \pm 0.001$ mHz, corresponding to a period of $684.9 \pm 0.5$ s. In addition to the main peak, several smaller peaks were observable. To identify whether these peaks were true signal or alias induced by observation gaps and orbital modulation, we fitted the light curve by a constant plus a sinusoidal function with a frequency fixed at 1.46 mHz to model the sinusoidal-like orbital modulation, and then subtracted the model from it. The residual light curve was analyzed again using the LS periodogram. The main peak from the orbital modulation was completely eliminated, but two significant peaks, which were evidently not the alias caused by the main peak and observation gaps, could be clearly observed in the power spectrum of residual (shown in Figure~\ref{sup_powspec}). The Gaussian fitting to these two peaks resulted in a frequency of $f_1 = 1.446 \pm 0.001$ mHz, corresponding to a period of $691.6 \pm 0.7$ s, which was close to the $693.5 \pm 1.3$ s superhump period proposed by~\citet{wan10}\footnote{Actually, in~\citet{wan10}, the superhump frequency was found to be $f_1 = 1.4442 \pm 0.0027$ mHz, whose corresponding period should be $692.4 \pm 1.3$ s instead of $693.5 \pm 1.3$ s. This revised period is consistent to what we found in this work.} for the most significant peak and $f_2 = 1.432 \pm 0.002$ mHz, whose corresponding period is $698.3 \pm 0.8$ s, for the other peak. We noticed that the relation between the frequency of the main peak and the frequencies of the two peaks from the residual light curve could be written as $f_n = f_0 - 0.014 \times n$ mHz where n=1 or 2. This indicates that the two peaks detected in the residual light curve power spectrum are the beat sidebands of a period of $0.8 \pm 0.1$ days modulation. We concluded that 4U 1820-30 is a superhump LMXB with accretion processing in a period of $\sim$0.8 days and the period of the most significant peak (i.e. $691.6 \pm 0.7$ s) is the superhump period, as the one detected in the FUV band~\citep{wan10}. Figure~\ref{sup_mod} shows the residual light curve folded by the superhump period where the peak-to-peak amplitude is approximately 1\%. Further discussions about the superhump in 4U 1820-30 are presented in section~\ref{sm}.

\section{Discussion} \label{dis}
\subsection{Orbital Period Change Rate}\label{obcr}

Using the phases obtained by the orbital modulations detected by the {\it NICER} 2017-2022 observations, we updated the orbital ephemeris of 4U 1820-30 by combining all historical records listed in Table 1 of~\citet{peu14}. The orbital period derivative evaluated from $\sim$46 years' baseline observations is $\dot P /P=(-5.21 \pm 0.13) \times 10^{-8}$ $yr^{-1}$, which is almost identical but more precise to the values reported by~\citet{van93b} and~\citet{peu14}, and similar to the one proposed by~\citet{chou01} whose deviation may be caused by phase jitter effect. No significant second order orbital derivative was detected with $|\ddot P|<5.48 \times 10^{-22}$ s $s^{-2}$ (2$\sigma$ upper limit). This indicates that the observed orbital period derivative of 4U 1820-30 was very stable, at least from 1976 to 2022. The observed negative $\dot P$ is incompatible to positive orbital period derivative ($\dot P/P > 8.8 \times 10^{-8}$ $yr^{-1}$) predicted by the standard scenario proposed by~\citet{rap87}. Although the orbital period derivative may be lower if other effects (e.g. tidal dissipation in the white dwarf driven by the eccentricity of binary system~\citep{pro12}) are considered, it should be still positive~\citep{peu14}.

The most popular scenario for explaining the discrepancy between the theoretical and observed values of the orbital period derivative is the acceleration of the binary system by the gravitation potential in the globular cluster NGC 6624~\citep{tan91,chou01,peu14}. In this scenario, the observed orbital period derivative can be express as  

\begin{equation}\label{orb_dev_sep}
{\dot P_{obs} \over P} \approx {\dot P_{int} \over P}+{a_c \over c}
\end{equation}

\noindent~\citep{phi92,per17} where ${\dot P_{obs}}$ is the observed orbital period derivative, ${\dot P_{int}}$ is the intrinsic orbital period derivative in the rest frame from the binary orbital evolution and $a_c$ is the line-of-sight acceleration from the cluster potential where $a_c<0$ means that the acceleration is towards the observer. Here, we ignore some other effects, such as the differential Galactic acceleration, acceleration towards the Galactic plane, and Shklovskii effect~\citep{shk70}, whose contributions to the observed orbital period derivative are only approximately $10^{-11}$ to $10^{-10}$ $yr^{-1}$~\citep{peu14,per17}, i.e. much smaller than the observed orbital derivative ($\sim 10^{-7}$ $yr^{-1}$). Our observation results show that the observed orbital period derivative is stable over $\sim$46 years, which implies that it is highly possible that both the intrinsic orbital period derivative and the external effect (e.g. acceleration in the globular cluster) are stable because it is very unlikely that the higher order terms of these two effects (i.e. $\ddot P_{int}/P$ and ${\dot a}_c/c$) have almost identical values but different signs to be canceled out to each other.

However, based on the cluster profile evaluated from the surface brightness and project radius of 4U 1820-30 to the center of NGC 6624,~\citet{peu14} found that the maximum acceleration of 4U 1820-30 was only $a_{c,max}/c \approx -4.6 \times 10^{-9}$ $yr^{-1}$, i.e. insufficient for explaining the observed orbital period derivative.~\citet{peu14} discussed several possibilities for further accelerate of this binary system in the globular cluster NGC 6624. The first is that  the binary system is accelerated by a flyby stellar size dark remnant, which can be a white dwarf, neutron star, or stellar size black hole. To preserve 4U 1820-30 as a triple system, which is currently the only way to explain the 171 days' periodic modulation, the mass of the dark remnant cannot exceed 6 $M_{\sun}$, otherwise the triple system will be destroyed due to the tidal force. Thus, a low mass dark remnant is preferred. On the other hand, the time span to maintain sufficient large acceleration to explain the binary's orbital period derivative would be shorter for lower mass dark remnant (20-160 years for 1-5.5 $M_{\sun}$, respectively), which makes it a  short-lived phenomenon. From the 2$\sigma$ upper limit of the second order orbital derivative, the variation time scale of $\dot P$, $|{\dot P}/{\ddot P}|$, would be longer than 65 years. However, if we consider the intrinsic orbital derivative, which is at least $8.8 \times 10^{-8}$ yr$^{-1}$, from Eq~\ref{orb_dev_sep}, the variation time scale of $a_c$ would be larger than 177 years under the assumption that ${\dot P}_{int}$ is a constant. Although this possibility cannot be completely excluded, the flyby dark remnant scenario appears unlikely to explain these observational results. In addition, it is scenario is very difficult to explain the spin period derivatives detected in the other three pulsars (PSR J1823-3021A,B,C)~\citep[see][Table 2]{peu14} in NGC 6624 because the probability of the these four sources experiencing similar flyby remnants simultaneously is very low~\citep{peu14}.

Another scenario discussed by~\citet{peu14} is that the binary system may be accelerated by an intermediate-mass black hole (IMBH) located at the center of the globular cluster. A mass of 7,500 $M_{\sun}$ IMBH is sufficient to explain the observed negative orbital period derivative with $a_c/c=a_{max}/c=-5.3 \times 10^{-8}$ yr$^{-1}$~\citep[see][Figure 7]{peu14}. Together with the other three pulsars, it needs a mass of 19,000 $M_{\sun}$ IMBH to explain their observed spin period derivatives. However, similar to the stellar dark remnant, such a large IMBH mass would destroy the triple system very quickly to make what we observed is a short-lived event. In addition,~\citet{peu14} proposed that a central concentration of dark remnants, such as white dwarfs, neutron stars and stellar size black holes, that sank into the center of NGC 6624 may be responsible for the negative orbital period derivative of 4U 1820-30, as well as the other three pulsars. They concluded that this scenario was preferable for explaining the observed period derivatives for the four sources. However, from the precise timing analysis of PSR J1823-3021A,~\citet{per17} found that this millisecond pulsar was very likely orbiting around an IMBH of at least 7500 $M_{\sun}$ mass located at the center of NGC 6624 with high eccentricity and they argued that it is not necessary to favor central dark remnant over an IMBH at center, but~\citet{gie18} constructed mass models without IMBH at the center that were able to explain the spin period derivatives of three millisecond pulsars.

However, for any of the scenario used to interpret the negative orbital period derivative or negative orbital phase drift, the internal orbital period derivative must be considered. Because the orbital modulation of 4U 1820-30 is likely a result of periodic occultation by the accretion stream of the vertical structure at the edge of the accretion disk~\citep{ste87,mor88,sam89,van93b},~\citet{van93b} suggested that the negative orbital phase drift of a -0.2 cycle detected from data collected between 1976 and 1991 may be due to accretion disk size change that causes the azimuth point of impact shift. Nevertheless, if the minimum intrinsic orbital period derivative, i.e. $8.8 \times 10^{-8}$ yr$^{-1}$~\citep{rap87}, taken into account, the azimuth point of impact has to shift 0.95 cycle, which makes this scenario unlikely~\citep{chou01}. A similar argument can also be applied to scenarios in which the observed negative orbital period derivative is caused by acceleration of the binary in the globular cluster. Considering the minimum intrinsic orbital period derivative, the acceleration along the line-of-sight of 4U 1820-30 should be at least $a_c/c=-1.41 \times 10^{-7}$ yr$^{-1}$.

Furthermore, large deviations of orbital period derivatives between the measured value and theoretical prediction, which considers orbital angular loss mechanisms only by gravitational radiation and magnetic braking, are often seen in LMXBs.~\citet{bur10} found that the observed orbital period derivative of the LMXB pulsar X 1822-371 was three order of magnitudes larger than the predicted value. The orbital period derivative of AX J1745.6-2901, which was first detected by~\citet{pon17}, was at least one order of magnitude  larger than the theoretical value. For the transient neutron star LMXB MXB 1659-298, the orbital period derivative was likely almost two order of magnitudes larger than the predicted value~\citep{iar18}. Similar large deviations have also been observed in accreting millisecond X-ray pulsars (AMXPs), such as SAX J1808.4-3658~\citep{dis08,bur09,san17}, IGR J17062-6143~\citep{bul21}, and SWIFT J1749.4-2807~\citep{san22}. The most interesting source is X 1916-053, an ultra-compact LMXB composed of a neutron star and a white dwarf, similar to 4U 1820-30, with an orbital period of 3000 s. An orbital period derivative of $\dot P/P=(1.62 \pm 0.34) \times 10^{-7}$ yr$^{-1}$, i.e. $\sim$200 times larger than the one induced by gravitational radiation, was reported by~\citet{hu08}. The most popular explanation for this large discrepancy is non-conservative mass transfer, i.e. a significant amount of mass transferred from the donor is ejected from the binary system instead of accreting onto the accretor. The outflow can carry away more orbital angular momentum to make the orbital period change faster than expected. For X 1916-053,~\citet{hu08} found that approximately 60\% - 90\% of the mass was transferred from the white dwarf companion outflowing from the binary system.~\citet{chou16} estimated that more than 60\% of the mass lost from the companion was ejected from X 1822-371. Moreover, the mass outflow ratios can be as high as more than 90\% for some LMXBs, such as IGR J17062-6143~\citep[90\%,][]{bul21}, MXB 1659-298~\citep[98\%,][]{iar18}, and SAX J1808.4-3658~\citep[99\%,][]{bur09}.~\citet{tav91} proposed a radiation-driven model in which the non-conservative mass transfer may be caused by the irradiation of the secondary and accretion disk by large a flux from the primary. It can induce a strong evaporative wind from the secondary and accretion disk that makes the orbital evolution time scale $\sim$100 times shorter than expected, such as X1916-053~\citep{hu08} and X 1822-371~\citep{chou16}. Alternatively, for AMXPs,~\citet{dis08} suggested that, similar to the mechanism  of black widow pulsars, the mass outflow can be driven by the pulsar wind, even in a quiescence state.

We suggest that the radiation-driven mass transfer mechanism, which induces non-conservative mass transfer, can also be applied to 4U 1820-30 that makes its intrinsic orbital period derivative much larger than the one that considers the orbital angular momentum lost only by gravitational radiation. Therefore, considering the orbital angular momentum lost by the outflow, the orbital period derivative, from Equation 3 in~\citet{dis08}, can be expressed as

\begin{equation}\label{dis08_3}
{\dot P_{orb} \over P_{orb}} =3\biggl\{{\biggl({\dot J_{orb} \over J_{orb}}\biggr)_{GR}}-{\dot M_2 \over M_2} \biggl[1-\beta q-\biggl(1-\beta \biggr)\biggl({{\alpha +q/3} \over {1+q} }\biggr) \biggr]\biggr\},
\end{equation}

\noindent where $(\dot J_{orb} / J_{orb})_{GR}$ is the orbital angular momentum loss driven by gravitational radiation, $M_1$ and $M_2$ are the masses of the neutron star and the companion, respectively, $q$ is the mass ratio ($q=M_2/M_1$), $\beta$ is the ratio of the mass accreting onto the neutron star, ${\dot M}_1=-\beta {\dot M}_2 $, and $\alpha$ is the specific angular momentum lost by the outflow expressed in units of the specific orbital angular momentum of the companion. For the Roche lobe filling companion, its radius $R_2$ is approximately equal to the Roche lobe radius $R_L=2/(3^{4/3})[1/(1+q)]^{1/3}$\citep{pac71}. Combined with the Kepler's 3rd law and the mass-radius relation of the companion, $R_2 \propto {M_2}^n$, from Eq~\ref{dis08_3}, we can have 

\begin{equation}\label{m2dot}
{\dot M_2 \over M_2} ={{\biggl({\dot J_{orb} \over J_{orb}}\biggr)_{GR}} \over {{5 \over 6} +{n \over 2} -\beta q -(1-\beta){{\alpha + q/3} \over {1+q}}}}
\end{equation}

\noindent and

\begin{equation}\label{pdot}
{\dot P_{orb} \over P_{orb}} ={({{3n} \over 2} - {1 \over 2})}{{\biggl({\dot J_{orb} \over J_{orb}}\biggr)_{GR}} \over {{5 \over 6} +{n \over 2} -\beta q -(1-\beta){{\alpha + q/3} \over {1+q}}}}
\end{equation}

\begin{figure}
\plotone{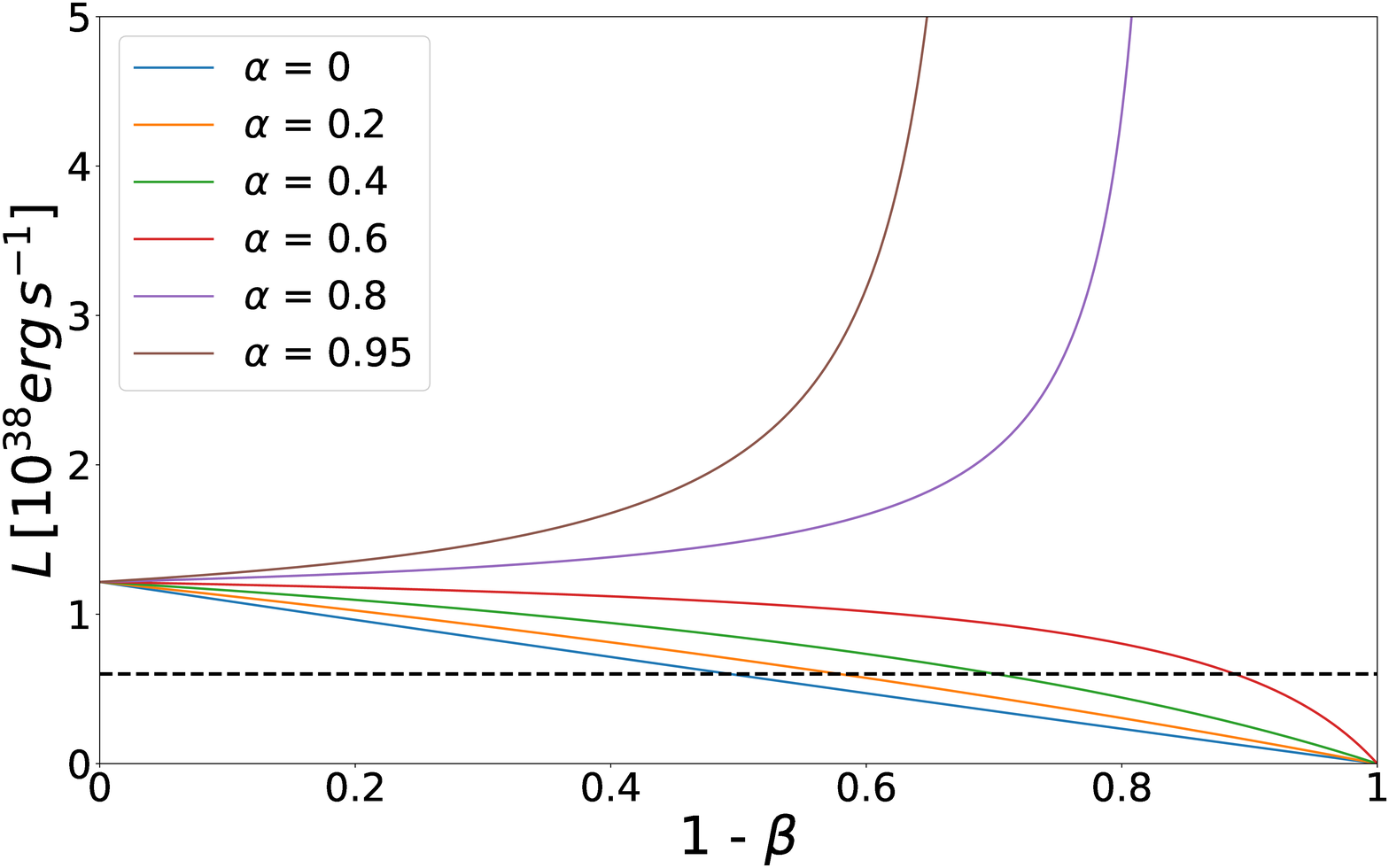}
\caption{Expected luminosity from Eq~\ref{lum} vs. mass outflow ratio (1-$\beta$) with different $\alpha$ values for $M_2 = 0.07 M_{\sun}$ . The dashed line is the observed luminosity $<L>=6.0 \times 10^{37}$ erg s$^{-1}$ \label{luminosity}}
\end{figure}

\begin{figure}
\plotone{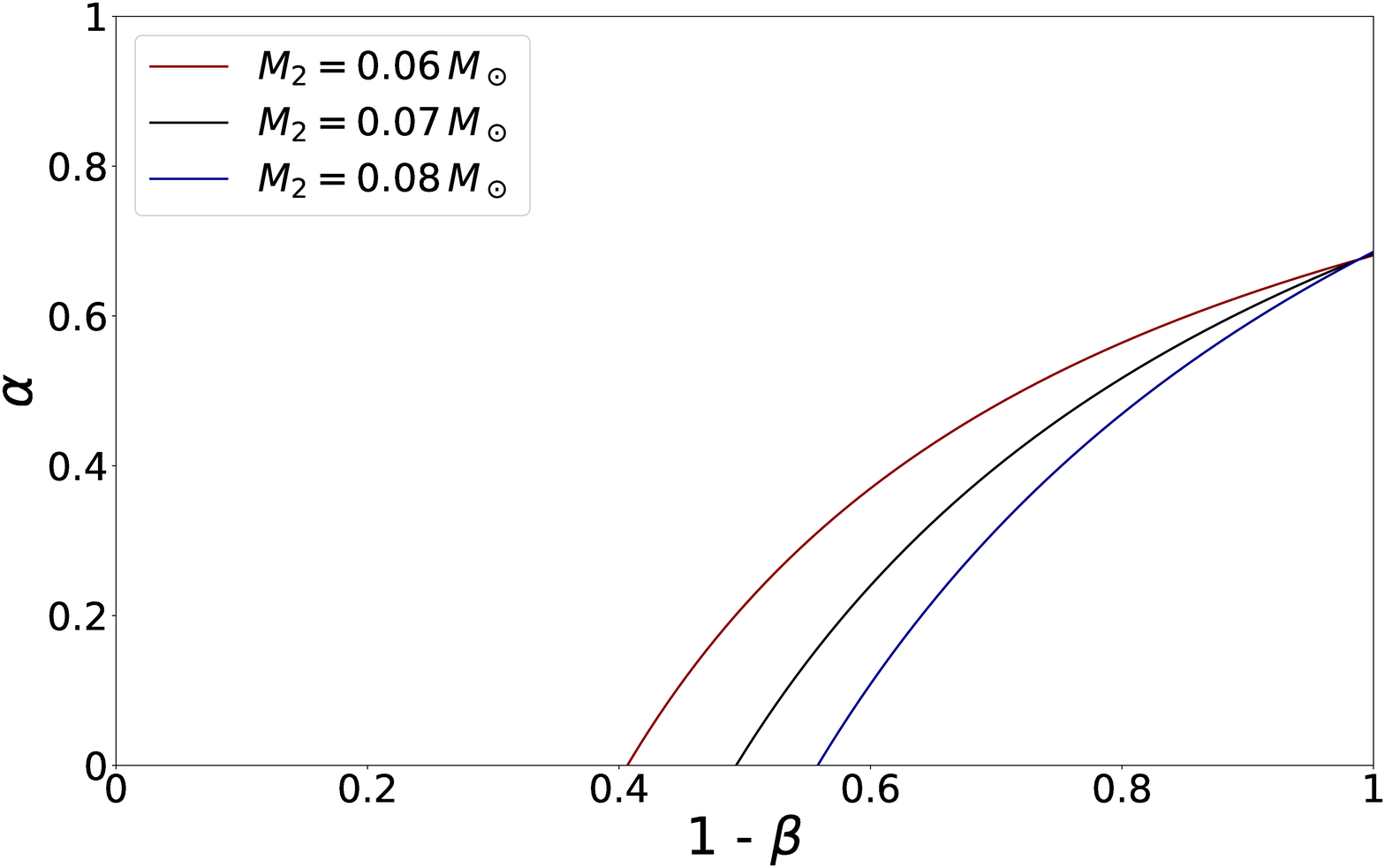}
\caption{Constraints of $\alpha$ and mass outflow ratio ($1-\beta$) for $L=6.0 \times 10^{37}$ erg s $^{-1}$ \label{alphabeta}}
\end{figure}

\begin{figure}
\plotone{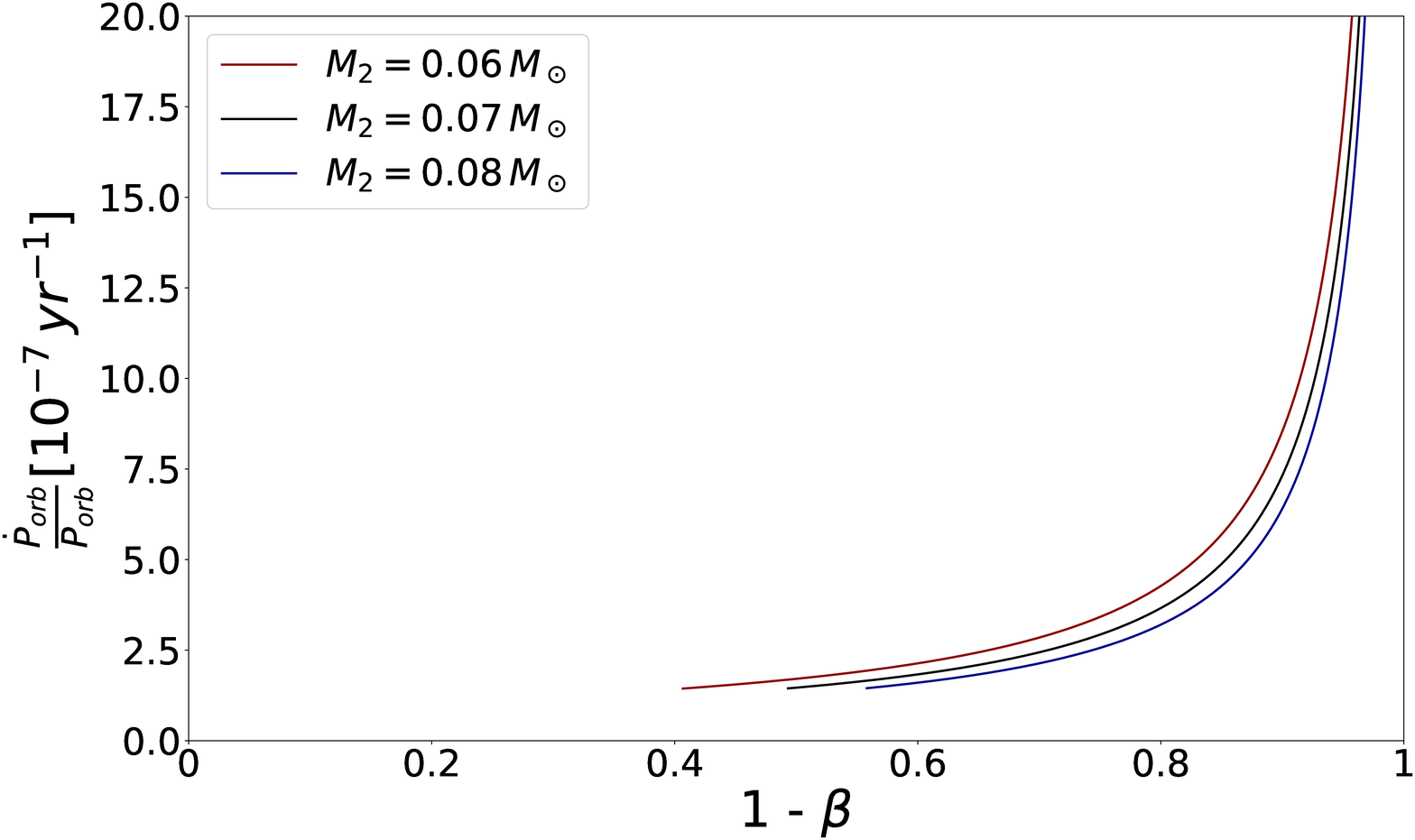}
\caption{Possible intrinsic orbital period derivative vs. mass outflow ($1-\beta$) for $L=6.0 \times 10^{37}$ erg s$^{-1}$ \label{possorbpder}}
\end{figure}

\noindent For 4U 1820-30, taking $M_{1}=1.4 M_{\sun}$, $M_{2}=0.07 M_{\sun}$, and $n=-1/3$ for the fully degenerate secondary, we obtained $\dot M_2 = -1.04 \times 10^{-8}$ $M_{\sun}$ yr$^{-1}$ for the mass conserved case ($\beta=1$). Thus we may evaluate the luminosity as $L= {{GM_{1} \dot M_{2}}/R_1}=1.22 \times 10^{38}$ erg s$^{-1}$ where $R_{1}=10^6$ cm is the neutron star radius that is larger than the mean observed value of $<L>=(6.0 \pm 0.2) \times  10^{37}$ erg s$^{-1}$~\citep{zdz07b}. If we assume that the deviation is caused by the non-conservative mass transfer, the luminosity can be evaluated as

\begin{equation}\label{lum}
L= -{{GM_{1} \beta \dot M_{2}} \over R_1} =-{{GM_{1} M_{2}} \over R_1}     {\beta{\biggl({\dot J_{orb} \over J_{orb}}\biggr)_{GR}} \over {{5 \over 6} +{n \over 2} -\beta q -(1-\beta){{\alpha + q/3} \over {1+q}}}}
\end{equation}

\noindent which enables us to constrain $\alpha$ and $\beta$ values. Figure~\ref{luminosity} shows the relation between the luminosity evaluated from Eq~\ref{lum} and mass outflow ratio (1-$\beta$). This indicates that the mass outflow ratio has a lower limit whereas $\alpha$ has an upper limit. From Eq~\ref{lum}, $\alpha$ can be written as a function of $1-\beta$, where $\alpha=f(1-\beta)$ is shown in Figure~\ref{alphabeta} for the companion's mass between 0.06$M_{\sun}$ and 0.08$M_{\sun}$. Evidently, at least $\sim 40\%$ (for $M_{2}=0.06M_{\sun}$) of the mass lost from the companion has to be ejected from the binary system. By applying the function $\alpha=f(1-\beta)$ to Eq~\ref{pdot}, orbital period derivative can be written as a function of mass loss ratio shown in Figure~\ref{possorbpder}. This implies that  the intrinsic orbital period derivative is no less than $1.44 \times 10^{-7}$ yr$^{-1}$ and can be very large for intense mass outflow.

According to the radiation-driven model proposed by~\citet{tav91}, the outflow can be ejected from the companion and accretion disk. If we assume the mass outflow is only ejected from the Roche-lobe filling companion, the minimum specific orbital angular momentum of the secondary star is at the inner Lagrangian point, which can be calculated using the formula $\alpha=\{1-[2/{(3^{4/3})q^{1/3}(1+q)^{2/3}}]\}^2$~\citep{dis08}. For 4U 1820-30, $\alpha=0.68$ when $M_{2}=0.07M_{\sun}$ and $M_{1}=1.4M_{\sun}$. This means that if the mass outflow is ejected from the companion only, we expect that $\alpha>0.68$. However, as shown in Figure~\ref{alphabeta}, we found that $\alpha < 0.68$, which implies that a part of the mass outflow is ejected from the accretion disk.

Although the observed negative orbital period derivative, $\dot P/P=(-5.21 \pm 0.13) \times 10^{-8}$ yr$^{-1}$, for 4U 1820-30 is very likely due to the acceleration of the binary system by the gravitational potential in the globular cluster NGC 6624~\citep{tan91,chou01,peu14}, when considering the acceleration value, the intrinsic orbital period derivative (i.e. $\dot P_{int}/P$ in Eq~\ref{orb_dev_sep}) cannot be neglected. From the above discussion, it is highly possible that 4U 1820-30 has significant mass outflow ratio ($>40\%$) and the intrinsic orbital period derivative is at least $1.44 \times 10^{-7}$ yr$^{-1}$ for $M_1 =1.4 M_{\sun}$ and $M_2 =0.6 M_{\sun}$, which makes the minimum line-of-sight acceleration equal to $|a_c/c|=1.96 \times 10^{-7}$ yr$^{-1}$. This value can be larger for a more intense mass outflow, as shown in Figure~\ref{possorbpder}. Therefore, we suggest that it is inappropriate to infer the acceleration of 4U 1820-30 in NGC 6624 using its observed orbital period derivative because there is too much uncertainty in it. In contrast to 4U 1820-30, the millisecond pulsars in NGC 6624 are likely better suited for inferring the accelleration in NGC 6624, as their  intrinsic spin period derivatives are relatively small ($\dot P/P \sim 10^{-10}$ yr$^{-1}$).

\subsection{Superhump Modulation in 4U 1820-30} \label{sm}

Superhumps are periodic modulations that are commonly observed in SU UMa type dwarf novae during their superoutburst states with periods 1\%-7\%  longer than the corresponding orbital periods \citep[see][for extensive review]{war95}. The superhump period is believed to be the beat period of the orbital period and accretion disk prograde precession period. If the mass ratio  $ q = M_2 / M_1 $ of a cataclysmic variable (CV) is less than 0.33, the accretion disk radius may extends to more than the 3:1 resonance radius. Under the influence of a tidal force from the companion, the accretion disk becomes asymmetric and starts precessing in the inertial frame~\citep{whi88,whi91}. In addition to the SU UMa systems, superhumps are also observed permanently (called permanent superhumps) in some CVs with no extreme brightness changes~\citep{pat99}. Negative superhumps (or infrahumps) have also been detected in permanent superhump systems, with periods a few percent shorter than the corresponding orbital periods due to disk nodal precession~\citep[e.g. AM CVn,][]{har98}.

In addition to the CVs, superhump modulations were observed in some LMXBs.~\citet{has01} indicated that the the 3:1 resonance condition ($q \lesssim 0.33$) could be more easily achieved in LMXBs than in CVs because of their larger accretors' masses. They proposed that superhumps might be detectable for LMXBs with orbital period less than 4.2 hours. For the black hole LMXBs, superhumps have been found in Nova Mus 1991, GRO JU0422+32~\citep{odo96}, XTE J1118+480~\citep{zur02}, and probably in GS 2000+25~\citep{odo96} and GRS 1915+105~\citep{nei07}. On the other hand, to explain the discrepancy between the periods observed in the X-ray band (3000 s) and optical band (3027 s) of dipping-bursting LMXB X1916-053,~\citet{whi89} suggested that the optical modulation is similar to the superhumps seen in SU UMa type dwarf novae due to disk precession in a period of $\sim$3.9 day whereas the X-ray dip period is the orbital period.~\citet{chou01a} detected a series of $\sim$3.9 day beat sidebands, including the optical period (3027 s), beside the X-ray dip period in the X-ray light curves collected by {\it RXTE} 1996 observations. This $\sim$3.9 day periodicity is likely due to the dip shape change caused by the slow disk precession in the inertial frame. Furthermore,~\citet{ret02} discovered a negative superhump with a period of 2979 s in an X-ray light curve and~\citet{hu08} found a 4.87 day periodic variation in the dip width from {\it RXTE} 1998 observations caused by the nodal precession of the accretion disk. These findings imply that X 1916-053 is similar to a permanent superhump system in the CVs~\citep{ret02,hu08}.   

The 693.5 s periodic modulation discovered in the FUV band is interpreted as a superhump in the 4U 1820-30 system~\citep{wan10}, that indicates that the accretion disk radius exceeds the 3:1 resonance radius and precesses with a period of $\sim$0.8 day. Although the superhumps in LMXBs are usually observed in longer wavelength bands, such as optical band for black hole LMXBs and X 1916-053, and FUV band for 4U 1820-30, the disk precesion effects are also observable in X-ray light curves. For instance, the X-ray dip shape changes with a $\sim$3.9 days disk apsidal precession ~\citep{chou01a} or $\sim$4.8 days disk nodal precession~\citep{hu08} in X 1916-053. Although 4U 1820-30 is not a dipping LMXBs, orbital modulation in the X-ray band is believed to be the result of absorption by the structure in the disk rim~\citep{ste87}. Therefore, as the disk undergoes processing, the beat sidebands beside the orbital peak, as detected by X 1916-053~\citep[see][Figure 5]{chou01a} with the period $P_{side}$ satisfied

\begin{equation}\label{side}
{1 \over {P_{side}}} = {1 \over {P_{orb}}} - {n \over {P_{prec}}}
\end{equation}

\noindent where $P_{orb}$ is the orbital period, $P_{prec}$ is the disk precession period, and $n=\pm 1, \pm 2,...$ in the power spectrum. In this work, we have detected two significant sidebands with periods of $P_1$=691.6$\pm$ 0.7 s (n=1 in Eq~\ref{side}), which was consistent with the proposed superhump period of 693.5$\pm$1.3 s~\citep{wan10}, and $P_2$=698.3$\pm$ 0.8 s (n=2). Combined with the orbital period detected in the power spectrum ($P_{orb}=684.9 \pm 0.5$ s), we estimated that the accretion disk is precessing in a period of 0.8$\pm$0.1 day.  

The mass ratio can be inferred from the orbital and superhump periods. From the periods we obtained in this work, the superhump excess is defined as $\varepsilon =(P_{sh}-P_{orb})/P_{orb}$, where $P_{sh}$ is the superhump period, i.e.  $\varepsilon$=0.010$\pm$0.001. According to the empirical equation $\varepsilon=0.18q+0.29q^2$ proposed by~\citet{pat05}, the mass ratio is estimated as $q$=0.05. On the other hand, for the low mass ratio,~\citet{goo06} derived a linear relation of the superhump excess and mass ratio, i.e. $\varepsilon=(0.2076 \pm 0.0003)q-4.1 \times 10^{-4}$. From this relation, the mass ratio is $0.049 \pm 0.005$ for 4U 1820-30. Both results indicate that the mass ratio is  $q \approx 0.05$. If the mass of the neutron star in 4U 1820-30 equals to its canonical value, i.e. 1.4$M_{\sun}$, the mass of the companion is 0.07$M_{\sun}$, which is consistent with the companion's mass range, i.e. 0.06$M_{\sun}$-0.08$M_{\sun}$, suggested by~\citet{rap87}.


Although no negative superhump signal is detected in the 4U 1820-30 {\it NICER} observations, based on the similarity between 4U 1820-30 and X 1916-053, we expect that 4U 1820-30 is also probably a permanent superhump system. If so, from the equation derived by~\citet{iar21}, the nodal precession frequency can be expressed as
\begin{equation}\label{nsh}
{\omega_n \over \omega_{orb}} = -{5 \over 32}   {q \over {1+q}} cos\delta
\end{equation}
\noindent where $\omega_n$ is the nodal precession frequency, $ \omega_{orb}$ is the orbital frequency, and $\delta$ is the tilt angle of the accretion disk. For a small tilt angle $cos\delta \approx 1$ and $q$=0.05 from the superhump period, the nodal precession period is evaluated as 1.07 days and the negative superhump period should be 679.9 s. The detection of negative superhump could further identify that 4U 1820-30 is a permanent superhump system as X 1916-053. However, the detection may be hampered by the $\sim$1 day observation gaps because it is close to the nodal precession period. Probably an uninterrupted 2-3 days observation is necessary to identify the existence of a negative superhump.

Conversely, even though our detection of orbital period sidebands with periods of $P_1$=691.6$\pm$0.7 s and $P_2$=698.5$\pm$ 0.8 s is possibly due to disk precession as the superhump detected in FUV band~\citep{wan10}, we cannot exclude the possibility that the sidebands are induced by a distant third star orbiting around the binary system. To explain the $\sim$171 days long-term modulation of the 4U 1820-30 system,~\citet{gri86,gri88} proposed that in the high-density star cluster core as NGC 6624, a star may be captured by the binary system and form a hierarchical triple system. As the third star orbiting the binary system, the eccentricity of the binary modulates with a period $P_{long}=KP^2_{outer}/P_{orb}$ where $P_{outer}$ is the third star orbital period, $P_{orb}$ is the binary orbital period, and $K$ is a constant of order unity, depending on the mass ratios and relative inclination of the third star orbit and binary~\citep{maz79}. The eccentricity variation changes the binary separation and thus the Roche lobe radius leads to mass loss rate modulation with a period of $\sim$171 days. This model was further confirmed by~\citet{zdz07a}. For 4U 1820-30, the triple model implies that the orbital period of third star is $\sim$1.1 days for $K \simeq 1$.~\citet{chou01} attempted to search for possible beat sidebands from the light curves collected by {\it RXTE} but failed, which was likely due to either the $\sim$1.1 day observation gaps or the variation being too small. Conversely, by applying the triple model, our detection of the beat sidebands indicates that the orbital period of third companion is 0.8$\pm$0.1 day, a little smaller than the expected $\sim$1.1 days. However, this value could be a factor of two larger or smaller owing to the different relative inclination angles~\citep{chou01}. The third companion may have various effects on the binary modulation with the third companion orbital period. As the binary system rotates, the beat sidebands from the power spectrum near the orbital signal may be observed. For instance, as the third companion star moving around the binary system in a prograde orbit, the small tidal force from it may slightly alter the binary separation. Because the Roche lobe radius is proportional to the binary separation and the mass loss rate is sensitive to the Roche lobe radius, it may cause a small variation of accretion stream from the $L_1$ point and change the vertical structure of the disk rim with a period of $\sim 692$ s, i.e. the beat period of binary and third star orbital period, thus the beat sidebands are observable in the X-ray light curves. However, to identify if the $\sim 692$ s periodicity is induced by the third companion, further evidence is required. For example,~\citet{chou01} considered the third companion with a maximum mass of 0.5$M_{\sun}$ and the motion of the binary radius relative to the mass center of the triple system was only $\sim 3.4 sin i_3$ lt-s where $i_3$ is the inclination of the third companion. Although this small motion only induces an orbital phase variation of less than 0.005, it is probably detectable by long-term continuous observations with a highly sensitive X-ray telescope.
 


\section{Summary} \label{sum}
4U 1820-30 is the most compact LMXB located near the center of globular cluster NGC 6624. In this work, we analyzed its orbital modulation from the X-ray light curves collected by {\it NICER} between 2017 to mid of 2022. In combination with the records since 1976, the orbital phase evolution in this 46 years can be well described by a quadratic curve with a detected orbital period derivative of $\dot P /P =(-5.21 \pm 0.13) \times 10^{-8}$ yr$^{-1}$. In addition to updating the orbital ephemeris, we found that no significant second order period derivative can be observed with a 2$\sigma$ upper limit of $|\ddot P|<5.48 \times 10^{-22}$ s s$^{-2}$.

The discrepancy between the detected (negative) and theoretically predicted (positive) orbital period derivative may be owing to the binary system accelerating in the gravitational potential in the globular cluster. However, as discussing the intrinsic orbital period derivative, we concluded that the 4U 1820-30 may have significant amount of mass outflow, which makes the intrinsic orbital derivative much larger than the one evaluated from the orbital angular momentum loss driven by gravitational radiation only. Therefore, it is difficult to infer the acceleration of the binary system by using the observed orbital period derivative because there is too much uncertainty in the intrinsic orbital period derivative.

On the other hand, from the X-ray light curves collected by {\it NICER}  between September 1 and 5 2017, a significant periodic modulation with a period of $691.6 \pm 0.7$ s, consistent with the superhump period discovered in the FUV band of the {\it Hubble Space Telescope}, was detected. This modulation is likely caused by the change of the absorption structure of the disk rim with a disk precession period of $0.8\pm0.1$ day coupling with the binary orbital motion. From the superhump period, the mass ratio is evaluated as $q \approx 0.05$. If the mass of neutron star is $1.4 M_{\sun}$, the companion's mass is 0.07$M_{\sun}$, which is consistent with the mass range 0.06 to 0.08$M_{\sun}$, suggested by~\citet{rap87}. While the $691.6 \pm 0.7$ s periodicity in X-ray band is likely the superhump period, it is still possible that the modulation is induced by a hierarchical third star orbiting around the binary system with an orbital period of $0.8\pm0.1$ day, such as the tidal force from the third companion changing the accretion stream, as well as the vertical structure of disk rim. Long-term observations with higher sensitivity X-ray telescope may be helpful for the identification.


\begin{acknowledgments}
The {\it NICER} data for this research were obtained from the High Energy Astrophysics Science Archive Research Center (HEASARC) online service, provided by the NASA/Goddard Space Flight Center. This work was supported by the National Science and Technology Council of Taiwan through the grant MOST 110-2112-M-008-029-.
\end{acknowledgments}

%

\vspace{5mm}
\facilities{ADS, HEASRAC, NICER}



\software{heasoft(v6.30), nicerdas(v009)}



\clearpage

\clearpage

\clearpage

\clearpage

\clearpage

\clearpage

\clearpage

\clearpage

\clearpage

\clearpage

\clearpage
\begin{table}
\begin{center}
\caption{ {\it NICER} Observation Grouping\label{nicergroup}}
\begin{tabular}{lccc}
\\
\tableline\tableline
ObsId  & Start Time & End Time & Exposure Time\\
       & (yy-mm-dd) &(yy-mm-dd)&   (s)      \\
\tableline
0050300108-109 & 17-06-26 & 17-06-27 & 6143 \\
\tableline
0050300113-117 & 17-07-09 & 17-07-13 & 5289 \\
\tableline
0050300119-120, 1050300101-102 & 17-07-14 & 17-07-18 &3664\\
\tableline
1050300105 & 17-08-22 & 17-08-22 & 4146 \\
\tableline
1050300107-108 & 17-08-28 & 17-08-29 & 33234 \\
\tableline
1050300109-110 & 17-08-30 & 17-08-31 & 28501 \\
\tableline
1050300111-112 & 17-09-01 & 17-09-02 & 26775\\
\tableline
1050300113 & 17-09-03 & 17-09-03 & 28161 \\
\tableline
1050300114 & 17-09-04 & 17-09-04 & 26849 \\
\tableline
1050300115 & 17-09-04 & 17-09-05 & 22491 \\
\tableline
1050300116-118 & 17-09-06 & 17-09-08 & 22161 \\
\tableline
1050300121-126 & 17-11-07 & 17-11-12 & 10635 \\
\tableline
1050300128-130 & 18-02-13 & 18-02-16 & 3231 \\
\tableline
1050300134-138 & 18-03-12 & 18-03-16 & 3416 \\
\tableline
2050300102-103 & 19-06-05 & 19-06-06 & 6677 \\
\tableline
2050300106-107 & 19-06-09 & 19-06-10 & 13820 \\
\tableline
2050300108-109 & 19-06-12 & 19-06-13 & 18960 \\
\tableline
2050300110-111 & 19-06-13 & 19-06-15 & 9868 \\
\tableline
2050300112     & 19-06-15 & 19-06-16 & 12539 \\
\tableline
2050300114-115 & 19-06-19 & 19-06-21 & 16008 \\
\tableline
2050300116-118 & 19-06-23 & 19-06-25 & 13903 \\
\tableline
2050300120-124 & 19-06-27 & 19-07-01 & 11050 \\
\tableline
2050300125-127 & 19-07-03 & 19-07-05 & 13085 \\
\tableline
2050300128-129 & 19-07-06 & 19-07-07 & 11671 \\
\tableline
3050300101-102 & 20-05-20 & 20-05-22 & 2630 \\
\tableline
3050300105-108 & 20-10-07 & 20-10-11 & 9975 \\
\tableline
3586010101-105 & 20-11-10 & 20-11-14 & 10185 \\
\tableline
4050300101  & 21-04-27 & 21-04-27 & 2928 \\
\tableline
4680010101-103 & 21-05-02 & 21-05-04 & 5583\\
\tableline
4050300104 & 21-05-19 & 21-05-19 & 4620 \\
\tableline
4663010101-102 & 21-06-19 & 21-06-20 & 19185\\
\tableline
4050300105-106 & 21-08-23 & 21-08-24 & 11590 \\
\tableline
4663020101-102 & 21-10-08 & 21-10-09 & 8270\\
\tableline
4680010105-106 & 21-10-13 & 21-10-14 & 2788\\
\tableline
4680010107-109 & 21-10-17 & 21-10-20 & 3824\\
\tableline
4680010112-115 & 21-10-24 & 21-10-27 & 12627\\
\tableline
4680010116 & 21-10-28 & 21-10-28 & 11937\\
\tableline
4680010117-119 & 21-10-29 & 21-10-31 & 11012\\
\tableline
4680010120-123 & 21-11-01 & 21-11-05 & 11783\\
\tableline
4680010124-128 & 21-11-06 & 21-11-11 & 5615\\
\tableline
5050300101-102 & 22-03-30 & 22-03-31 & 4465 \\
\tableline
5050300107-111 & 22-04-04 & 22-04-09 & 8663 \\
\tableline
5604010103,5050300112-116 & 22-04-16 & 22-04-20 & 8595\\
\tableline
5604010301-303 & 22-04-30 & 22-05-02 & 9453\\
\tableline
5604010401-402 & 22-06-12 & 22-06-13 & 10535\\
\tableline
5604010501-503 & 22-06-26 & 22-06-28 & 5993\\
\tableline
5604010601-602 & 22-07-08 & 22-07-09 & 610349\\
\tableline
\end{tabular}
\end{center}
\end{table}

\clearpage
\begin{table}
\begin{center}
\caption{Average Orbital Phase of 4U 1820-30 from 2017-2022 {\it NICER} Observations\label{nicerphase}}
\begin{tabular}{lcc}
\\
\tableline\tableline
Mean Arrival Time  & Phase & Error\\
($BJD_{TDB}$) &  & \\
\tableline
2457999.65 & -1.58 & 0.03\\
2458178.08 & -1.76 & 0.05\\
2458656.07 & -1.74 & 0.05\\
2459078.61 & -1.86 & 0.07\\
2459429.82 & -1.96 & 0.11\\
2459719.61 & -2.03 & 0.03\\
\tableline
\end{tabular}
\end{center}
\end{table}

\end{document}